\documentclass[conference]{IEEEtran}
\usepackage{amsmath,amsfonts}
\usepackage{algorithm}
\usepackage[noend]{algpseudocode}
\usepackage{algpseudocode}
\usepackage{array}
\usepackage[normalem]{ulem}
\usepackage{subfigure}
\usepackage{textcomp}
\usepackage{stfloats}
\usepackage{url}
\usepackage{verbatim}
\usepackage{graphicx}
\usepackage{cite}
\usepackage[dvipsnames]{xcolor}
\usepackage{float}

\usepackage{hyperref}
\hypersetup{
    colorlinks=true,
    linkcolor=blue,
    filecolor=black,      
    urlcolor=blue,
    citecolor=blue,
}

\usepackage[left=1.62cm,right=1.62cm,top=1.9cm]{geometry}
\newcommand{\cmmnt}[1]{}
\begin{document}

\title{Multi-UAV Multi-RIS QoS-Aware Aerial Communication Systems using DRL and PSO}   
\author{
    \IEEEauthorblockN{Marwan Dhuheir\IEEEauthorrefmark{1},
    Aiman Erbad\IEEEauthorrefmark{1}, Ala Al-Fuqaha\IEEEauthorrefmark{1}, and
    Mohsen Guizani\IEEEauthorrefmark{2}
    }
    \IEEEauthorblockA{\IEEEauthorrefmark{1}Division of Information and Computing Technology, College of Science and Engineering,\\ Hamad Bin Khalifa University, Qatar Foundation, Doha, Qatar.
    }
    \IEEEauthorblockA{\IEEEauthorrefmark{2}Machine Learning Department, Mohamed Bin Zayed University of Artificial Intelligence (MBZUAI), Abu Dhabi, UAE.
   }
}
\maketitle

\begin{abstract}
Recently, Unmanned Aerial Vehicles (UAVs) have attracted the attention of researchers in academia and industry for providing wireless services to ground users in diverse scenarios like festivals, large sporting events, natural and man-made disasters due to their advantages in terms of versatility and maneuverability. However, the limited resources of UAVs (e.g., energy budget and different service requirements) can pose challenges for adopting UAVs for such applications. Our system model considers a UAV swarm that navigates an area, providing wireless communication to ground users with RIS support to improve the coverage of the UAVs. In this work, we introduce an optimization model with the aim of maximizing the throughput and UAVs coverage through optimal path planning of UAVs and multi-RIS phase configurations. The formulated optimization is challenging to solve using standard linear programming techniques, limiting its applicability in real-time decision-making. Therefore, we introduce a two-step solution using deep reinforcement learning and particle swarm optimization. We conduct extensive simulations and compare our approach to two competitive solutions presented in the recent literature. Our simulation results demonstrate that our adopted approach is 20 \% better than the brute-force approach and 30\% better than the baseline solution in terms of QoS.
\end{abstract}


\section{introduction}
Recently, unmanned aerial vehicles (UAVs) have been utilized as wireless communication base stations (UAV-BS) due to their exceptional advantages, including mobility, efficient cost, quick setup, maneuverability, etc., in contrast to traditional techniques \cite{padro2019comparison}. UAVs can be used as a rapid solution for providing wireless communication in multiple mission-critical scenarios such as festivals, large sporting events, and scenarios involving infrastructure failures due to natural or man-made disasters \cite{dhuheir2024meta,9456851}. To cover the service area, one UAV might not be sufficient to provide efficient services to all user elements (UE) on the ground; hence, multi-UAVs have emerged as an appropriate solution \cite{10002339}. Nonetheless, various challenges need to be addressed for the effective use of UAVs including path planning, energy consumption, and resource allocation.

In the past decade, Reﬂecting Intelligent Surfaces (RIS) attracted the attention of researchers to improve the signal of communication links. RIS is an effective technique for improving the quality of communication links, especially in overcrowded areas where this signal is interrupted by several obstacles (e.g., buildings, etc.). Generally, RIS contains an array of passive reflecting elements that help reflect the signal by adjusting its phase shift. By adjusting the phase shift, the throughput of the data transmission can be significantly enhanced. Furthermore, the array elements are passive; hence, it is more energy efficient and cost-effective because it requires less hardware than the relaying techniques in traditional communication systems \cite{10118908, 9804341}. 

Several studies have been conducted to improve data throughput using UAVs and RIS. Li et al. \cite{8959174} investigated UAV trajectory planning with passive beamforming of the RIS to maximize throughput. Liu et al. \cite{9615225} studied the data rate maximization by incorporating RIS for UAV downlink transmissions. All the aforementioned studies investigated the single RIS to improve the throughput; however, multi-RIS studies with UAV swarms have not been investigated in the recent literature. Muhammad et al. \cite{9804341} investigated multi-RIS in the case of using MEC as a processing center, which is ineffective in providing services in mission-critical scenarios like disaster scenarios. In fact, the joint use of multi-RIS with UAV swarms to improve the throughput and coverage has not been investigated in the literature, which is the key contribution of this article.

The main contribution of this article is to investigate the joint use of multi-RIS multi-UAV swarms to serve UE gatherings in a power law distribution where some areas are busier than others, like festivals, large sporting events, or disaster scenarios. Deployed RIS is used to improve the QoS of UEs that require different services, including video, data, and audio. The main contribution of this article can be summarized as follows:
\begin{itemize}
    \item To maximize the throughput and coverage, multi-objective optimization is formulated while considering pivotal constraints, including UAV energy, minimum SNR required for different services, and hardware consideration of RIS phase shift.
    \item To tackle the above optimization problem, deep reinforcement learning (DRL) is adopted to find the UAVs' path planning to maximize the coverage of UAV swarms. To maximize the throughput, Particle Swarm Optimization (PSO) is adopted to find the optimal configuration of RIS phase shift. These two solutions are generated iteratively until the optimal solution is obtained. 
    \item Extensive simulations were performed to assess the performance of the proposed solutions. Moreover, two competitive solutions are implemented as baselines for the adopted solution, one using a brute-force approach and the other using recent work from the literature.
\end{itemize}

The rest of this article is organized as follows: Section \ref{info_system_model} describes the system model. In Section \ref{problem_formulation}, we delineate the problem formulation. Section \ref{proposed_solution} presents the details of the proposed approach. Section \ref{Performance_evaluation} presents the performance results of the proposed approach and compares it to other approaches from the recent literature. Finally, Section \ref{conclusion} concludes our study and discusses future research directions.

\section{system model}
\label{info_system_model}

\begin{figure}[!ht]
    \centering
    \includegraphics[width=0.35\textwidth]{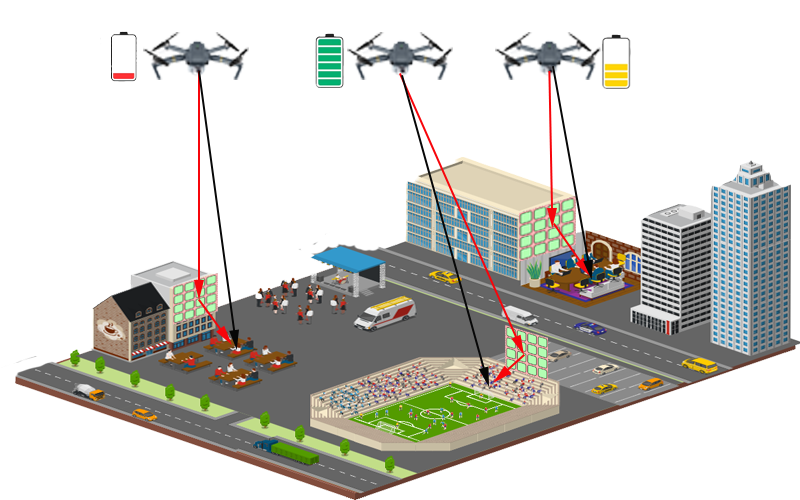}
    \caption{System Model for UAV swarm covering a service area while focusing on regions with higher user densities (i.e., strategic locations). The swarm's mission is data collection from ground devices.}
    \label{System_Model}
\end{figure}

The system model, as depicted in Figure \ref{System_Model}, outlines the key elements involved in our research problem. In this model, the geographical service area is partitioned into uniformly sized cells, with each UAV being allocated to a specific cell. These UAVs are tasked with delivering wireless communication services to ground-based users. In this scenario, a swarm consisting of $U$ UAVs is responsible for traversing the service area to provide wireless communication services to Ground User Elements (UEs). The wireless communication needs to cover a wide range of areas, accommodating a diverse range of UE devices with varying service requirements, including video, data, and audio. The UAV swarm, denoted by $\bar{U} = \{1, 2, \ldots, U\}$, is tasked with meeting the service requirements of the set of UEs, denoted by $\bar{N} = \{1, 2, \ldots, N\}$, where $N$ represents the maximum number of UEs on the ground. It is important to note that the positions of the UEs $O_i = [x_i, y_i, 0]$, where $i \in \bar{N}$ are assumed to be readily available through the utilization of the Global Positioning System (GPS) or similar systems. 

The UAV swarm navigates the area at time frame $T$ where $T > 0$ in seconds (s). The time frame $T$ is divided into time intervals $t$, where $0 \leq t \leq T$. Consequently, for each UAV $u$ belonging to the UAV swarm denoted as $\Bar{U}$, we denote the 3D positions at time $t$ by $O_u^t = [x_u^t, y_u^t, z_u^t] \in \mathbb{R}^3$. 
For simplicity and to maintain generality, we assume that the UAVs maintain a constant altitude, denoted as $z_u$, which adheres to regulatory and safety requirements. These UAVs strategically navigate the designated area, focusing on regions with higher user density (i.e., strategic locations). This approach aims to efficiently provide a range of services tailored to the diverse needs of users while ensuring that their Quality of Service (QoS) requirements are met. This strategy serves the dual purpose of maximizing the number of served users and delivering a satisfactory user experience. To improve the communication link between UAVs and UEs, an $r \in \bar{R}$ RIS equipped with $M$ reflecting elements is adopted, where $\bar{R} = \{1, \dots, R$\}, and $R$ is the maximum number of RIS. The locations of RIS are assumed to be known at $O_r^t = [x_r^t, y_r^t, z_r^t]$ as shown in Figure \ref{System_Model}. 

\subsection{Wireless Channel Model}
In this section, we introduce the wireless communication model employed between the UAVs, multi-RIS, and UEs. In our approach, we assume that wireless communication is conducted by using FDMA. Also, we note that the communication link between the UAVs and the UEs is referred to as the direct link, while the communication link from a UAV to an RIS that terminates at a UE is referred to as an indirect link.
\subsubsection{Direct Link}
In our approach, and for practical considerations, the obstacle information, including their number, height, and locations, might not be known; hence, the randomness of the availability of line-of-sight (LoS) and non-line-of-sight (NLoS) channels of the air-to-ground link between UAVs and UEs are considered. 
Hence, the probability of LoS expression is given by \cite{mozaffari2017mobile}:
\begin{equation}
\small 
    P_{u,i}^{LoS} = \frac{1}{1+\omega_1 \exp{-\omega_2 [\theta_{u,i}-\omega_1]}}
    \label{Los_equation}
\end{equation}
where $\omega_1$ and $\omega_2$ are constant parameters and their values are specified based on the type of the environment, $\theta_{u,i}$ represents the elevation angle between the UAV $u$ and the UE device $i$. Particularly, $\theta = \frac{180}{\pi} \times \sin^{-1}(\frac{z_u}{d_{u,i}})$, where $d_{u,i} = \sqrt{(x_u- x_i)^2 +(y_u - y_i)^2 + z_u^2}$ is the distance between the UAV $u$ and UE $i$. The probability of NLoS is given by, $P_{u,i}^{NLoS} = 1 - P_{u,i}^{LoS}$.

Thus, the channel gain of the UAV and UE in direct link can be expressed as follows \cite{9934086}:
\begin{equation}
\small
    h_{u,i}^t = (P_{u,i}^{LoS}) (d_{u,i})^{-\alpha_1} + (P_{u,i}^{NLoS}) (\alpha_2) (d_{u,i})^{-\alpha_1}
\end{equation}
where $\alpha_1$ represents the path loss exponent, and $\alpha_2$ denotes the attenuation factor for NLoS.

\subsubsection{Indirect Link}
We assume that the elements of RIS are following a linear array distribution. Also, the communication link between UAV and RIS, and RIS to UE follows a Rician fading channel in which it experiences small-scale fading. Hence, the communication link between UAV $u$ and RIS $r$, $h_{u,r}^t \in \mathbb{C}^{M \times 1}$ can formulated as follows \cite{9934086}:

\begin{equation}
\small
    h_{u,r}^t = \sqrt{\mu(d_{u,r})^{-\alpha}} \sqrt{\frac{K}{K+1}} \bar{h}_{u,r}^t
\end{equation}
where $\mu$ denotes the average path loss power gain at reference distance $d_0$ = 1, $d_{u,r}$ denotes the distance between UAV $u$ and RIS $r$, where $d_{u,r} = \sqrt{(x_u- x_r)^2 +(y_u - y_r)^2 + (z_u-z_r)^2}$, $K$ represents the Rician factor, and $\Bar{h}_{u,r}$ represents the LoS array components of RIS, which can be computed by:
\begin{equation}
\small
    \bar{h}_{u,r}^t = \underbrace{\Biggr[1, e^{-j\frac{2\pi}{\lambda}\tau\phi_{u,r}^t}, \dots , e^{-j\frac{2\pi}{\lambda}(M-1)\tau\phi_{u,r}^t} \Biggr]^T}_{\text{array components}}
\end{equation}
where $\phi_{u,r}^t = \frac{x_u - x_r}{d_{u,r}}$ is the cosine of the angle of the signal traversing from RIS to UAV, $\tau$ represents the separation between RIS elements, and $\lambda$ represents the wavelength of the carrier signal.

Likewise, the channel gain between the RIS and UE can be computed, which $h_{r,i}^t \in \mathbb{C}^{M \times 1}$ by:
\begin{equation}
\small
    h_{r,i}^t = \sqrt{\mu(d_{r,i})^{-\alpha}} \sqrt{\frac{K}{K+1}} \bar{h}_{r,i}^t
\end{equation}
where $d_{r,i}$ is the distance between RIS and UE where $d_{r,i} = \sqrt{(x_r- x_i)^2 +(y_r - y_i)^2 + h_r^2}$, and $\phi_{r,i}^t = \frac{x_r - x_i}{d_{r,i}}$
\begin{equation}
\small
    \bar{h}_{r,i}^t = \underbrace{\Biggr[1, e^{-j\frac{2\pi}{\lambda}\tau\phi_{r,i}^t}, \dots , e^{-j\frac{2\pi}{\lambda}(M-1)\tau\phi_{r,i}^t} \Biggr]^T}_{\text{array components}}
\end{equation}

Let $\Theta^t$ denote the phase shift matrix of the RIS in the $t-th$ time slot where $\Theta^t = diag\{e^{j\Phi_1^t}, \dots, e^{j\Phi_M^t}\}$, $\Phi^t$ represents the phase shift of the $m-th$ array component at time slot $t$, and $m = \{1, \dots, M$\}. For hardware limitations, the phase shift is chosen from a vector of specific values that follow the hardware constraints. In particular, the set of RIS elements array is given as $\Phi^t \in \psi = \{0, \frac{2\pi}{W}, \dots, \frac{2\pi (W-1)}{W}$, where $W = 2^b$ and $b$ is the number of bits that identifies the components of phase shift of RIS elements array. Thus, the signal to noise ratio $\Gamma$ is given as follows: 
\begin{equation}
\small
    \Gamma^t_{u,r,i} = \frac{P|h_{u,i}^t+h_{r,u}^{t,H}\Theta^t h_{r,i}^t|^2}{\sigma^2}, \quad \forall u \in \Bar{U}, \forall r \in \Bar{R}, \forall i \in \Bar{N}
\end{equation}
where $P$ is the transmit power, and $H$ represents the operator of conjugate transpose.

In noisy environments, the users are served when the wireless transmission is above a threshold, i.e., the i-th user is set to be covered with its services if the probabilistic mean of $\Gamma_{u,r,i}^{t}$ exceeds a predefined $\Gamma_{th}^{j}$ (dB) threshold as follows:
\begin{equation}
\small
    \Gamma_{u,r,i}^t \geq \Gamma_{th}^{j}, \quad \forall u \in \Bar{U}, \forall r \in \Bar{R}, \forall i \in \Bar{N},
    \label{SNR_requiremnts}
\end{equation}
where $j \in \{1, 2, 3\}$,  $j = 1$ indicates the set of all users requesting video, $j = 2$ indicates the set of all users requesting data, and $j = 3$ indicates the set of all users requesting audio.

Hence, the data rate that is delivered to the UE can be calculated as follows:
\begin{equation}
\small
    \rho_{u,r,i}^t = B \log_2(1+\Gamma_{u,r,i}^t), \quad \forall u \in \Bar{U}, \forall r \in \Bar{R}, \forall i \in \Bar{N},
    \label{data_rate}
\end{equation}
where $B$ is the allocated bandwidth.

\subsection{UAVs Energy Consumption Model}
UAVs are energy-limited due to their restricted onboard battery capacity. The battery's lifespan is influenced by various factors, such as the UAV's energy source, its type, weight, and speed. Usually, the UAV's energy usage can be categorized into two main components: propulsion energy and communication energy. Communication energy is significantly smaller in scale compared to propulsion energy and, as a result, it is excluded from the energy model in our system. To model the propulsion energy, we utilize the propulsion-power model designed for rotary-wing UAVs as in \cite{9513250}:

\begin{equation}
\begin{aligned}
\small
    \varepsilon_{prop,u} = \underbrace{\eta_i \sqrt{\Bigg( \sqrt{\Big(1+\frac{v_u^4}{4v_0^4}}\Big)-\frac{v_u^2}{2v_0^2}\Bigg)} }_\textbf{Induced Power} +
    \underbrace{\eta_b \Big(1+\frac{3v_u^2}{v_{tip}^2}\Big) }_\textbf{Blade Power} \\ + \underbrace{\frac{1}{2}f_0\varphi r D_a v_u^3}_\textbf{Parasite Power}
    \end{aligned}
    \label{prop_energy}
\end{equation}

where $\eta_i$ refers to the blade profile power and $\eta_b$ refers to the induced power, $v_{tip}^2$ indicates the speed of the UAV's rotor blade, $v_0$ is the rotor induced velocity, $f_0$ refers to fuselage drag ratio, $r$ refers to the rotor solidity, $\varphi$ refers to the air density, and $D_a$ is the area of the rotor disc. To calculate the hovering power consumption, equation (\ref{prop_energy}) is used with zero speed of the UAV, i.e., $v_u = 0$, as follows:
\begin{equation}
\small
    \varepsilon_{hov,u} = \eta_i + \eta_b
\end{equation}
Therefore, the total energy consumption of UAV $u$ at time slot $t$ is obtained as follows:
\begin{equation}
\small
    \varepsilon_{u,tot}^t =   
        \left\{ 
        \begin{array}{l}
            \varepsilon_{prop,u} \times t \;\text{\;if $v_u > 0$}\\
            \varepsilon_{hov,u} \times t \; \text{\;if $v_u = 0$}\\
        \end{array}
        \right. 
\end{equation}

Due to their limited battery lifespan, UAVs need to have sufficient energy to accomplish their mission; hence, we add a constraint to ensure sufficient energy is available for the UAVs during their mission. The battery status $\Omega_u^t$ at each time slot $t$ can be obtained as follows:
\begin{equation}
\small
    \Omega_u^t = \Omega_u^{t-1} - \varepsilon_{u,tot}^t
\end{equation}
where $\Omega_u^{t-1}$ is the battery level at the end of $t-1$. Let $\Omega_u^{0}$ denote the battery capacity before the mission starts, in which $\Omega_u^{0} = \Omega_u^{init} + \Omega_u^{min}$, where $\Omega_u^{init}$ is the battery capacity of the UAV that is assigned for the mission, and $\Omega_u^{min}$ is the minimum battery level for the UAV to return to its central station, therefore, $\Omega_u^t \in [\Omega_u^{min}, \Omega_u^{0}]$

\section{problem formulation}
\label{problem_formulation}
In this section, we present the formulation of our multi-objective optimization problem. 
At each time step, the UAVs aim to maximize the number of served users, $C_{u,i}$ by planning UAV paths to prioritize serving the cells with more users. This optimization is designed considering several critical constraints, including the battery levels of UAVs, the minimum SNR required to ensure that the UEs receive their essential services, and the minimum safe distance to prevent collisions with other UAVs. Thus, the overall optimization can be expressed as follows:
\begin{equation}
\small
    \max_{x,y,\Theta} \sum_{u \in \Bar{U}}
    \sum_{i \in \Bar{N}}
    \sum_{r \in \Bar{R}}
    (\rho_{u,r,i}^t, C_{u,i})
    \label{objective_fun}
\end{equation}
    \begin{equation}\tag{\ref{objective_fun}a}
    \small
    \text{subject to} \qquad
         \Omega_u^t \geq \Omega_u^{min}, \quad \forall u \in \Bar{U}, \forall t \in T
         \label{C1}
    \end{equation}
    \begin{equation}\tag{\ref{objective_fun}b}
    \small
         d_{u,k} \geq 2D_{max}, \quad \forall u, k \in \Bar{U}
         \label{C2}
    \end{equation}
    \begin{equation}\tag{\ref{objective_fun}c}
    \small
         \Gamma^t{u,r,i} \geq \Gamma^{t}{th}, \quad \forall u \in \Bar{U}, \forall r \in \Bar{R}, \forall i \in \Bar{N}, \forall j \in \{1,2,3\}
         \label{C3}
    \end{equation}
    \begin{equation}\tag{\ref{objective_fun}d}
    \small
         \Phi^t_r \in \psi, \quad \forall t \in T, \forall r \in \Bar{R}
         \label{C4}
    \end{equation}
    \begin{equation}\tag{\ref{objective_fun}e}
    \small
       0 \leq \Phi^t_m \leq 2\pi , \quad \forall t \in T, \forall m \in M
         \label{C5}
    \end{equation}

The multi-objective function in equation (\ref{objective_fun}) aims to maximize the number of covered users and throughput of transmission data in the case of heterogeneous requirements of users. The ground users have different service requirements, and the UAVs need to optimize their positions and the phase of the RIS elements to satisfy their demand and improve their QoS. 

Constraint \ref{C1} is set to ensure that UAVs have sufficient energy to accomplish their mission of providing wireless communication services to the users gathering in a condensed area like a stadium or a festival. Constraint \ref{C2} refers to the safe distance between UAVs to avoid collisions and a threshold of $2D_{max}$ are met while they are in motion. Constraint \ref{C3} refers to the minimum SNR of each user based on its requirement. This constraint is set to ensure successful communications between UAVs and ground users. Constraint \ref{C4} is set to ensure that the values of phase shifts are within their possible range, and \ref{C5} is set to ensure that the value of RIS phase shift is within allowable hardware constraints.

The multi-objective optimization problem in equation \ref{objective_fun} is nonconvex due to the nonlinearity in equations \ref{objective_fun}, \ref{C1}, \ref{C2}, \ref{C3} and \ref{C4}. Hence, obtaining a linear programming solution using traditional optimization techniques is difficult. We propose a less complex and dynamic solution using deep reinforcement learning (DRL) and particle swarm optimization (PSO) algorithms. The following section delineates the details of the proposed solution.

\section{the proposed solution}
\label{proposed_solution}
In this section, we devise a low-complexity solution to solve the problem formulated in the previous section. Our proposed solution utilizes DRL to solve the problem of UAV path planning and PSO to solve the problem of optimal configuration of RIS phase shift elements. DRL and PSO work iteratively to find the optimal positions of UAVs that maximize the number of served users and the optimal phase shift that maximizes the throughput of the communication to ground users. The solution of the multi-objective function in \ref{objective_fun} can be described as follows:

\subsection{Coverage Maximization:}
To solve the problem of coverage maximization, a dynamic and real-time solution is adopted using DRL. The DRL agent interacts with the environment to learn the optimal policy. The optimization problem of coverage maximization and its related constraints can be expressed as follows:
    \begin{equation}
    \begin{aligned}
    \small
    \textbf{P1: } \max_{x,y} \sum_{u \in \Bar{U}}
    \sum_{i \in \Bar{N}}
    ( C_{u,i})
    \label{P1}
    \end{aligned}
    \end{equation}
    \begin{equation}\tag{\ref{P1}a}
    \small
     \text{subject to} \qquad
        \ref{C1}, \ref{C2}
        \label{P1_constraints}
    \end{equation}

We formulate the \textbf{P1} problem in \ref{P1} into MDP by defining its five crucial parameters $(S, A, \gamma, R, P)$, where $S$ refers to the state of the environment, $A$ indicates the action of the agent that leads to the optimal knowledge of the environment,  $\gamma$ refers to the discount factor which indicates the effect of the agent on making a specific decision on the future relative actions, where its value is in the interval of $0 \leq \gamma < 1$, $P$ indicates the transition probability of the following state given the current state and its related actions $P(S^{t+1}|s^t,a^t)$, and $\forall \; S^{t+1},s^t \in S, a^t \in A$. $R: S \times A \rightarrow \mathbb{R}$ indicates the reward function that the agent receives from making a specific action $a^t$ on the state $s^t$ and producing $s^{t+1}$, where $r^t = r(s^t, a^t, s^{t+1})$.
 \subsubsection{State}
 The state is a pivotal parameter in learning the optimal policy in which the agent depends on it to understand the environment and increase the reward function. The set of states at each time step $t$ includes the following details:
 \begin{equation}
 \small
     s^t = [x_u^t, y_u^t, P_u^*, P_i^*, U_u^t]
 \end{equation}
 where $x_u^t, y_u^t$ indicate the position of the current UAV at time step $t$, $P_u^*$ refers to the positions of all UAVs within the grid, and their already covered users in the time step $t$, $P_i^*$ refers to the positions of all UE in the ground, and $U_u^t$ indicates the index of the current UAV.

\subsubsection{Action}
The action $a^t$ of the agent at each time step $t$ is the UAV next positions $(x_u, y_u)$, which includes nine directions, and the agent gets a positive reward if the agent’s action maximizes the number of covered users which is compatible with the objective function of the problem in \ref{P1}.
\subsubsection{Reward function}
In order to improve the performance of the system model, the reward function needs to be well set to encourage the agent to learn the optimal policy. The reward at each time step $r^t = 1$ if the agent chooses the path that maximizes the coverage and respects the constraints in \ref{P1_constraints}, otherwise $r^t = 0$.

Algorithm \ref{path_planning_algorithm} delineates the steps for UAV path planning using a DRL solution to maximize the coverage of ground users while respecting the constraints of UAVs' minimum battery level and avoiding collision with other UAVs in the grid. According to this algorithm, an agent receives a reward if it chooses a path that maximizes the coverage, avoids collision, and has a minimum battery level to accomplish the mission (lines: 7-8).
\begin{algorithm}
\caption{DRL for UAVs Path Planning}
\label{path_planning_algorithm}
\begin{algorithmic}[1]
\small
\State \textbf{Inputs: } \text{U,N, $\gamma, \epsilon$}
\State \textbf{Output: } \text{UAVs path planning.}
\For{each episode $t$ $\in$ $T$}
\For{each UAV $u \in \{1,2,3,...,U\}$}
    \State \text{$s^t = [x_u^t, y_u^t, P_u^*, P_i^*, U_u^t]$}
    \State \text{choose action $a^t$ based on $\epsilon$}
    \If{constraints in equation(\ref{P1}) and (\ref{P1_constraints})} 
    \State \text{$R_t = R_t + 1 \qquad$}
    \EndIf
    \State \text{observe $S^{t+1}$}
    \State \text{observe $R^t$}
    \State \text{update system with UAVs new  locations}
    \State \text{save $(S^t, A^t,r^t,S^{t+1})$ in replay memory}
    \State \text{sample a minibatch of $(S_i, A^t,r^t,S^{t+1})$}
    \State \text{$\theta_{old} \gets \theta$}
    \State \text{$error = R^t + \gamma V(S^{t+1}) - V(S^t)$}
\EndFor
\EndFor
\end{algorithmic}
\end{algorithm}
\subsection{Throughput Maximization:}
The second sub-problem is to find the optimal configuration of the RIS phase shifts in order to maximize the throughput of wireless communication. The optimization problem of throughput maximization and its related constraints can be expressed as follows:
    \begin{equation}
    \small
    \begin{aligned}
    \textbf{P2: } \max_{\Theta}
    \sum_{u \in \Bar{U}}
    \sum_{i \in \Bar{N}}
    \sum_{r \in \Bar{R}}
    (\rho_{u,r,i}^t)
    \label{P2}
    \end{aligned}
    \end{equation}
    \begin{equation}\tag{\ref{P2}a}
    \small
    \text{subject to} \qquad
        \ref{C3}, \ref{C4}, \ref{C5}
        \label{P2_constraints}
    \end{equation}

PSO is a metaheuristic optimization algorithm inspired by bird predation behavior, where individuals in a group share information to collectively improve problem-solving. It involves updating the coordinates of particles, symbolizing potential solutions, in each iteration to move towards the best location and particle, refining the overall solution. This collaborative process transitions from disorder to order, leading to the acquisition of a feasible solution. Algorithm \ref{RIS_phase_shift} delineates the steps of PSO to find the optimal configuration of RIS phase shifts to maximize the throughput. The process of checking the possible solution continues until convergence or the maximum number of iterations is reached (lines: 5-11), then based on the number of population, the velocity $V(i)$ is calculated and then the next $\Theta$ is updated before checking the main objective function of problem \ref{P2} for the maximum value of the throughput (lines: 6-8). The new value of RIS phase shift $\Theta$ is compared with the maximum value and then the outcome will update the value of the best RIS Phase shift $\Theta$ (lines: 10-11).

\begin{algorithm}
\caption{PSO for RIS phase shift}
\label{RIS_phase_shift}
\begin{algorithmic}[1]
\small
\State \text{$\chi = 1 - \frac{1}{c} + \frac{\sqrt{|c^2-4c|}}{2}$, $c = c1 + c1$} and \text{$J_1,J_2 \in \mathbb{C^{M\times M}}$}
\State \textbf{initialization: } \text{choose $\Theta_{best}$ and $\Theta(i)$ randomly}
\State \textbf{Inputs: } \text{iteration maximum ($itrmax$), population size ($npop$)}
\State \textbf{Output: } \text{RIS phase shift ($\Theta$)}
\For{$t$ $\leftarrow$ $itrmax$}
\For{$i$ $\leftarrow$ $npop$}
 \State $V(i+1) = \chi ( V(i) + c1 J_1 \oplus (\Theta_{best}-\Theta(i)))$ \text{$\qquad \qquad \qquad \qquad \qquad \qquad + c2 J_2 \oplus (\Theta_{best}-\Theta(i))$}
 \State $\Theta(i+1) = \Theta(i) + V(i+1)$
 \State \text{$\Theta_{costfunc} = $ Evaluate equation (\ref{data_rate})}
 \If{$\Theta_{costfunc} > \Theta_{best}$}
    \State \text{$\Theta_{best} = \Theta_{costfunc}$}
 \EndIf
\EndFor
\EndFor
\end{algorithmic}
\end{algorithm}

 
\section{simulation results and analysis}
\label{Performance_evaluation}
This section presents the performance evaluation of our system model. For our simulation, we used a 400 $\times$ 400 m area with 400 UEs distributed according to a power law distribution, i.e., some areas are busy, and some areas are not, resembling people gathering for a festival and in a stadium. We applied threshold values of 30 dB, 25 dB, and 20 dB for different service requirements, namely video, data, and audio, respectively. For environment type, we used urban area with parameters of $\omega_1$ and $\omega_2$ as 11.95 and 0.14, respectively. The bandwidth value is 1 MHz, carrier frequency is 1 GHz, AWGN noise power $\sigma^2$ is -170dBm, discount $\gamma$ is 0.95, and the learning rate $\alpha$ is 0.0005. 
We compared our adopted solution with the metaheuristic solution using a brute-force algorithm and a baseline; namely random waypoint (RWP) mobility  \cite{8671460}. The brute-force algorithm employs an exhaustive search among the available solutions, selecting the one with the highest outcome, while the baseline solution tests a random selection of UEs-UAVs within the area\cite{8671460}. We emphasize that due to the complexity of exploring all possible choices in the brute-force solution when calculating the RIS phase shift, which needs to adhere to constraint \ref{C5} in \textbf{P2}, we conducted an exhaustive search among 1000 possible values.

\begin{figure}[!ht]
    \centering  \includegraphics[width=0.32\textwidth]{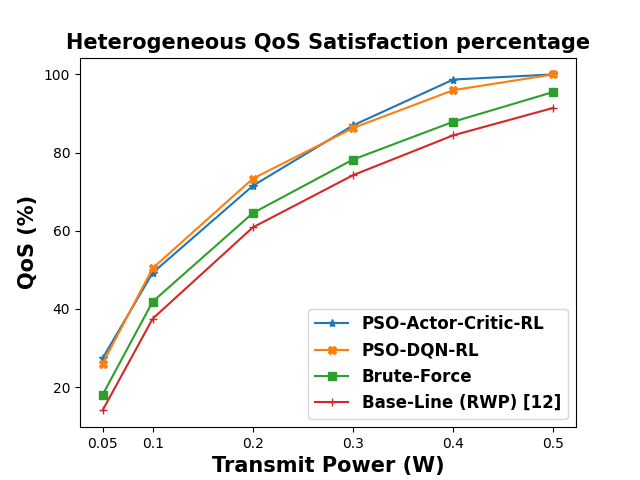}
    \caption{The percentage of satisfying the QoS of UE when increasing the UAVs transmit power and comparing our adopted solution with two competitive solutions.}
    \label{fig:transmit_power}
\end{figure}

Figure \ref{fig:transmit_power} presents the simulation results of the percentage of service satisfaction for heterogeneous UEs as we increase the UAVs' transmit power. We tested our adopted solution using two DRL algorithms, actor-critic and DQN, to solve the first objective function and the PSO algorithm to address the second objective function. As the UAVs' transmit power increases, the percentage of QoS improves, allowing the UAVs to allocate more resources to enhance throughput for heterogeneous UEs. Additionally, our adopted solution achieves superior QoS in comparison to the two competitive solutions. However, the brute-force approach outperforms the baseline solution, albeit at the expense of more computations, since it explores numerous possible solutions and selects the one that maximizes QoS.

\begin{figure}[!ht]
    \centering  \includegraphics[width=0.32\textwidth]{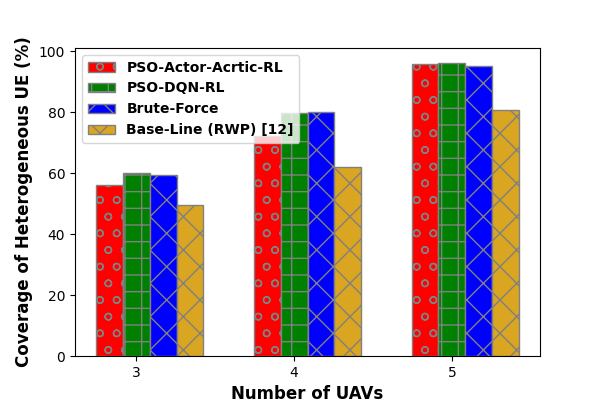}
    \caption{The coverage percentage of UE when increasing the number of UAVs and comparing our adopted solution with two competitive solutions which represents the outcomes of objective \textbf{P1}. }
    \label{fig:served_users}
\end{figure}

Figure \ref{fig:served_users} presents the coverage percentage as the number of UAVs increases. As shown in the figure, as the number of UAVs increases, the percentage of covered UEs also increases. Our adopted solution, utilizing PSO-DRL, achieves comparable results to the brute-force algorithm since the brute-force method checks the entire domain of UAV paths to identify the best paths for maximizing the coverage percentage. In contrast, the baseline solution is the least effective, as the UAVs move randomly within the grid. 
\begin{figure}[!ht]
    \centering  \includegraphics[width=0.32\textwidth]{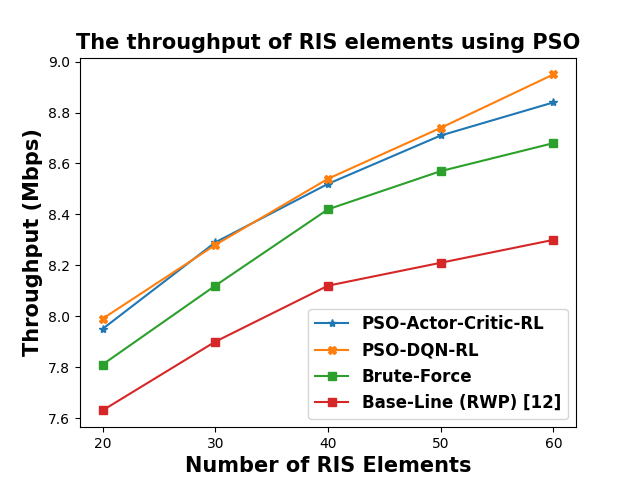}
    \caption{The throughput evaluation when increasing the RIS elements in PSO algorithm and comparing it with two competitive solutions which represents the outcome of the objective function in \textbf{P2}.}
    \label{fig:throughput}
\end{figure}

Figure \ref{fig:throughput} presents the results of throughput when increasing the number of RIS elements $M$. As the number of RIS elements is increased, the throughput of data transmission increases. Our adopted solution using PSO-DRL achieves the best among the two competitive solutions using brute-force and baseline solutions. Moreover, actor-critic and DQN-based DRL approaches achieve comparable results, whereas the baseline solution achieves the least throughput among the solutions.
\section{conclusion}
\label{conclusion}
In this article, we investigated the optimization of UAV paths and RIS phase shift configurations. We formulated the problem as a multi-objective optimization problem that seeks to maximize throughput and coverage. We propose a two-step solution using DRL and PSO. We compared our proposed solution with two competitive solutions to illustrate the effectiveness of our approach in providing better QoS, coverage, and throughout. In future work, we plan to study the interference and the Doppler effects on UAV movements.

\section*{Acknowledgment}
This work was made possible by NPRP grant \# NPRP13S-0205-200265 from the Qatar National Research Fund (a member of Qatar Foundation). The findings achieved herein are solely the responsibility of the authors.
\bibliographystyle{IEEEtran}
\bibliography{bibliography.bib}
\end{document}